\documentclass[preprintnumbers,amsmath,amssymb,floatfix,12pt]{revtex4}

\usepackage{graphicx}
\usepackage{dcolumn}
\usepackage{bm}

\begin{document}

%

\let\a=\alpha      \let\b=\beta       \let\c=\chi        \let\d=\delta
\let\e=\varepsilon \let\f=\varphi     \let\g=\gamma      \let\h=\eta
\let\k=\kappa      \let\l=\lambda     \let\m=\mu
\let\o=\omega      \let\r=\varrho     \let\s=\sigma
\let\t=\tau        \let\th=\vartheta  \let\y=\upsilon    \let\x=\xi
\let\z=\zeta       \let\io=\iota      \let\vp=\varpi     \let\ro=\rho
\let\ph=\phi       \let\ep=\epsilon   \let\te=\theta
\let\n=\nu
\let\D=\Delta   \let\F=\Phi    \let\G=\Gamma  \let\L=\Lambda
\let\O=\Omega   \let\P=\Pi     \let\Ps=\Psi   \let\Si=\Sigma
\let\Th=\Theta  \let\X=\Xi     \let\Y=\Upsilon

%

%

\def\cA{{\cal A}}                \def\cB{{\cal B}}
\def\cC{{\cal C}}                \def\cD{{\cal D}}
\def\cE{{\cal E}}                \def\cF{{\cal F}}
\def\cG{{\cal G}}                \def\cH{{\cal H}}
\def\cI{{\cal I}}                \def\cJ{{\cal J}}
\def\cK{{\cal K}}                \def\cL{{\cal L}}
\def\cM{{\cal M}}                \def\cN{{\cal N}}
\def\cO{{\cal O}}                \def\cP{{\cal P}}
\def\cQ{{\cal Q}}                \def\cR{{\cal R}}
\def\cS{{\cal S}}                \def\cT{{\cal T}}
\def\cU{{\cal U}}                \def\cV{{\cal V}}
\def\cW{{\cal W}}                \def\cX{{\cal X}}
\def\cY{{\cal Y}}                \def\cZ{{\cal Z}}

%

\newcommand{\Ns}{N\hspace{-4.7mm}\not\hspace{2.7mm}}
\newcommand{\qs}{q\hspace{-3.7mm}\not\hspace{3.4mm}}
\newcommand{\ps}{p\hspace{-3.3mm}\not\hspace{1.2mm}}
\newcommand{\ks}{k\hspace{-3.3mm}\not\hspace{1.2mm}}
\newcommand{\des}{\partial\hspace{-4.mm}\not\hspace{2.5mm}}
\newcommand{\desco}{D\hspace{-4mm}\not\hspace{2mm}}



\title{\boldmath On non-minimal coupling of the inflaton}

\author{Namit Mahajan
}
\email{nmahajan@prl.res.in}
\affiliation{
 Theoretical Physics Division, Physical Research Laboratory, Navrangpura, Ahmedabad
380 009, India
}


\begin{abstract}
The non-minimal coupling of the inflaton to gravity, $\xi R\phi^2$, is known to alleviate
the smallness (fine-tuning) of the quartic coupling $\lambda$ in the chaotic
inflation with $\phi^4$ potential. A large $\xi$ is required to obtain the correct CMB power spectrum
while we find that a small value $\sim 1/6$ seems to be preferred from spectral index. There are issues related to
conformal transformations, choice of frame and natural value(s) of $\xi$ for a given potential. We revisit some of these
issues and invoke certain field theoretic arguments in order to address the same. A rather strong and general
conclusion reached, based on the requirements of renormalizability and finiteness of specific matrix elements in a quantum theory, 
is that it is generically not possible to eliminate the non-minimal coupling by going from the Jordan to the Einstein frame via conformal transformations.
We also comment on Higgs inflation.
\end{abstract}

\maketitle
The standard inflationary scenario (see \cite{Lyth:1998xn} for a review) provides a good description of the observed
universe, including the cosmic microwave background (CMB) spectrum and growth of structure. For chaotic inflation with a quartic potential, $V(\phi) = \lambda \phi^4$, (with inflation taking place around the typical grand unified theory (GUT) scale) requires $\lambda \sim {\mathrm few}\times 10^{-13}$. This value seems significantly smaller than the natural value expected within a reasonable
particle physics scenario. For example, in the standard model (SM), the Higgs quartic coupling $\sim 0.1$ at the weak scale. This essentially is at the heart of the problem. In all what has been said above, the coupling of the matter (scalar field here) to gravity is minimal. It was recognised long ago \cite{Fakir:1990eg} that allowing for non-minimal coupling of the inflaton with the gravity sector can soften the problem related to abysmally small value of $\lambda$ (see also \cite{Komatsu:1999mt}). Specifically, the action now takes the form
\begin{equation}
 S^{(J)} = \int d^4x \sqrt{-g}\left\{\frac{1}{2}(\frac{1}{\kappa^2} - \xi \phi^2)R + \frac{1}{2}(\partial\phi)^2 - V(\phi) \right\} \label{jordanaction}
\end{equation}
where $\kappa^2 = 8\pi G$, $R$ is the Ricci scalar, $\phi$ is a single component real scalar field and $V(\phi) = \lambda \phi^4$. The superscript $(J)$ refers to the Jordan frame and reflects fact that the gravity sector is no longer of the Einstein-Hilbert form i.e. $\sqrt{-g} R$ form. $\xi = 0$ leads to the minimal coupling while $\xi=1/6$ is referred to as conformal coupling since the classical action possesses conformal invariance. At this point it is instructive to mention that in the Higgs inflation models \cite{Bezrukov:2007ep}, the Higgs potential is assumed to be a quartic one since at large field values ($\phi >> v$, with $v$ being the Higgs vev), the quadratic term can be safely neglected. All the discussion will be centered around a quartic potential. Most of the aruments made below are expecetd to qualitatively apply to other potentials as well, though some care must be exercised.

A conformal transformation $g_{\mu\nu}\to \hat{g}_{\mu\nu} = \Omega^2(x)g_{\mu\nu}$ could be made with a suitable $\Omega$ to bring the above Jordan frame action to the corresponding Einstein frame action such that there is no non-minimal term (for a review see \cite{Faraoni:1998qx}). Below all the hatted quantities refer to Einstein frame. In the present case, identifying 
\begin{equation}
 \Omega^2(x) = \Omega[\phi(x)] = 1 - \kappa^2 \xi\phi^2
\end{equation}
 and redefining the metric gets rid of the non-minimal term altogether. There is a price to pay, however. The scalar field is no longer canonically normalised. This could be fixed by redefining the scalar field as well:
\begin{equation}
 d\hat{\phi} = \frac{[1- \kappa^2\xi(1-6\xi)\phi^2]^{1/2}}{\Omega^2}d\phi
\end{equation}
With this redefinition of the metric and the scalar field, the action in the Einstein frame reads
\begin{equation}
 S^{(E)} = \int d^4\hat{x}\sqrt{-\hat{g}}\left\{\frac{1}{2}(\frac{1}{\kappa^2}\hat{R} + \frac{1}{2}(\partial\hat{\phi})^2 )- \hat{V}(\hat{\phi}) \right\} \label{einsteinaction}
\end{equation}
where the potential now takes the form
\begin{equation}
\hat{V}(\hat{\phi}) = \frac{V(\phi)}{(1-\kappa^2\xi\phi^2)^2}
\end{equation}
As is evident from Eq.(\ref{einsteinaction}), in this frame, $\hat{\xi} = 0$ though the potential is no longer in the polynomial form. In particular, for the quartic potential and also Higgs inflation (where again the potential in the Jordan frame is approximated to the quartic form), the potential obtained after the conformal transformations (expressed in terms of the redefined field variable) is very flat if $\xi >> 1$. This then leads to successful inflation.

The actions in the Jordan and the Einstein frame are related to each other by variable/field redefinitions and therefore are expected to be equivalent. This equivalence is only up to the additive boundary terms \cite{Saltas:2010ga}, and care must be exercised in dealing with these, particularly when going beyond the tree level results. In particular, they are expected to yield identical results for any physical quantity like the CMB power spectrum, spectral index ($n_s$).  In fact, it can be shown that the gauge invariant curvature perturbation
in the Einstein frame coincides  with that in the Jordan frame i.e. $\hat{\zeta}(\hat{x}) = \zeta(x)$ \cite{Komatsu:1999mt}, \cite{Makino:1991sg}. The only missing piece in establishing the equivalence of the physical quantities in the two frames at the quantum level is to show that the two vacua are identical. At least in the perturbative sense, it can be shown that for interacting theories, 
$\vert\hat{0}\rangle = \vert 0\rangle$ \cite{Kubota:2011re}. Therefore, the power spectrum calculated in either frame is the same, given by (assuming for simplicity a scale invariant spectrum and not showing insignificant dependence on the scale which is irrelevant for the present discussion)
\begin{equation}
{\mathcal{P}}_s \sim \Delta^2_{\zeta} \sim \frac{\lambda}{\xi^2}\frac{N^2}{72\pi^2} \label{powerspectrum}
\end{equation}
where $N \sim 60$ is the relevant number of e-folds and $\Delta^2_{\zeta} = 2.4\times 10^{-9}$ \cite{Hinshaw:2012fq}.
 Further, appropriately transforming the slow roll parameters, it is easily seen that $\hat{n}_s = n_s$. Following \cite{Komatsu:1999mt}, 
\begin{equation}
\hat{n}_s = 1 - 2\hat{\epsilon} - 2\hat{\alpha} \label{ns}
\end{equation}
where
\begin{equation}
\hat{\epsilon} \sim \frac{1}{\kappa^2}\left(\frac{\hat{V}_{,\hat{\phi}}}{\hat{V}}\right)^2  = 1.5\times 10^{-4}\frac{1+6\xi}{6\xi}\left(\frac{70}{N(\hat{t}_k)}\right)^2
\end{equation}
\begin{equation}
\hat{\alpha} \sim \frac{1}{\kappa^2}\left[\left(\frac{\hat{V}_{,\hat{\phi}}}{\hat{V}}\right)^2 - \frac{\hat{V}_{,\hat{\phi}\hat{\phi}}}{\hat{V}}\right]
= 1.4\times 10^{-2}\left(\frac{70}{N(\hat{t}_k)}\right) + 3\times 10^{-4}\frac{1+6\xi}{6\xi}\left(\frac{70}{N(\hat{t}_k)}\right)^2
\end{equation}

From Eq.(\ref{powerspectrum}) it is clear that for the natural expected values of $\lambda \sim {\mathcal{O}}(10^{-2}-10^{-1})$, $\xi \sim few\times 10^{4}$ is needed. Following a similar line of reasoning, \cite{Bezrukov:2007ep} found $\xi_{{\mathcal{P}}_s} \sim 44700\sqrt{\lambda}$ in the context of Higgs inflation model, assuming $\xi>>1$. What is important is to note that in this regime of $\xi$ values, the constraints from the CMB power spectrum imply $\xi_{{\mathcal{P}}_s} \propto \sqrt{\lambda}$. On the other hand, one could make use of Eq.(\ref{ns}) to obtain a constraint on $\xi$. This may have a certain advantage over the CMB power spectrum constraint since in the slow roll approximation ( at least to the leading order in slow roll parameters), all the quantities entering the relation are ratios of the derivatives of the potential and the potential itself, thereby canceling the dependence on the coupling constant. Plugging $n_s = 0.971$ \cite{Hinshaw:2012fq}, one finds $\xi_{n_s} \sim 0.13$, which is not too different from $\xi = 1/6 \sim 0.17$. 
However, at this point what is important to note is the fact that the two values of $\xi$ obtained above are in total contradiction to each other. This, we believe is well known to experts but is often not stated explicitly.

A priori $\xi$ can take any value and it would be good if based on some physical (and/or mathematical) requirements and consistency, a range of values for $\xi$ could be identified (see \cite{Faraoni:1998qx}, \cite{Faraoni:1996rf}, \cite{Faraoni:2000wk} for a clear discussion on various possibilities and restrictions; see also \cite{Basak:2012pc}). In a quantum field theory setup, a non-zero $\xi$ appears or is rather forced upon us due to quantum corrections even if the classical action has $\xi=0$. It was shown \cite{Callan:1970ze} that an improvement term is needed in the energy momentum tensor, and the precise form of the improvement term coincides with the non-minimal term in question. Collins argued further \cite{Collins:1976vm} that the softness and finiteness of the matrix elements of $\Theta_{\mu\nu}$ and $\Theta^{\mu}_{\mu}$ (in the context of $\phi^4$ theory) imply $\xi = \frac{1}{4}\frac{n-2}{n-1}$, where $n$ is the number of space-time dimensions. For $n=4$, one obtains $\xi = 1/6$. It is worth mentioning that there is perhaps no unique way to fix the exact form and parameter dependence of the improvement term and its coefficient \cite{Joglekar:1988dv}, and one has to proceed on a case by case basis depending on the relevant interactions the scalar field has. Further, 
within the realm of metric gravity theories, and when the scalar field does not have a gravitational origin, it is known that the equivalence principle forces $\xi$ to be $1/6$ (see \cite{Faraoni:1996rf}). All these observations strongly pointing to $\xi = 1/6$ in the context of $\phi^4$ theory seem far from being a mere coincidence. $\xi=1/6$ for a pure $\phi^4$ potential is known not to lead to inflationary solutions (see \cite{Faraoni:1996rf}, \cite{Faraoni:2000wk} for a clear discussion), though such solutions do exist if $m^2\phi^2$ and/or cosmological constant terms are also taken into consideration. One could therefore have these terms present in the
action but for all practical purposes they can be safely neglected, leading to quasi-conformal invariance of the action, at least in the case when the field values are way bigger than the mass parameter(s) in the theory. 

If the above observations are to be taken at the face value, then there is an apparent puzzle: the disparate values of the non-minimal coupling parameter $\xi$ coming from CMB power spectrum and spectral index. Assuming a small value of $\xi$ as suggested by above considerations would imply the return of the problem of small $\lambda$. Let us focus on the Higgs inflation model within the context of SM \cite{Bezrukov:2007ep} for concreteness. As mentioned above, $V(\Phi) \propto \lambda (\Phi^{\dag}\Phi - v^2)^2 \longrightarrow \lambda \Phi^{\dag}\Phi$ when $\vert\Phi\vert^2>>v^2$, which simplifies further by working in the unitary gauge where the potential now resembles the quartic coupling of a real scalar field. Transforming to the Einstein frame and assuming $\xi>>1$, the potential takes the form $\hat{V}(\hat{\phi}) = \frac{\lambda M_P^4}{4\xi^2}\left(1-e^{-\frac{2\hat{\phi}}{\sqrt{6}M_P}}\right)$. Finally imposing the power spectrum constraint, one obtains $\xi \sim 44700\sqrt{\lambda}$. The large value of $\xi$ is obtained assuming $\lambda \sim {\mathcal{O}}(0.1)$, as is relevant for Higgs physics. It is known that in a quantum field theory the couplings and masses all run with the energy scale, and within SM, the quartic coupling $\lambda$ runs in such a way that for the central values of the parameters like top quark mass and the strong coupling constant, $\lambda$ becomes negative (vacuum stability problem \cite{Sher:1988mj}) at scales smaller than the GUT scale. This would imply that Higgs inflation in the context of pure SM would be rather difficult to achieve. Moreover, it has been argued that tree level unitarity implies that the theory is valid only till the scales $<<M_P$ \cite{Barbon:2009ya}. However, the quartic coupling around that scale, even if one chooses the SM parameters to avoid negative values of $\lambda$ at such a scale, flows to a rather small value, leading to an apparent conflict. The way around is to realise that $\xi$ also runs and satisfies its own renormalization group equation (RGE). It is the ratio $\lambda/\xi^2$ that is needed to be held at a small value consistent with $\Delta^2_{\zeta}$. Theoretical consistency requires adding to the Lagrangain (in the Jordan frame) terms corresponding to the external gravitational field (beyond the $R$ term already taken into account \cite{Odintsov:1993rt}):
\begin{equation}
 {\mathcal{L}}_G = \Lambda + a_1R^2 + a_2C^2_{\mu\nu\alpha\beta} + a_3{\mathcal{G}} + a_4\Box R
\end{equation}
where $\Lambda$ is the cosmological constant, $C_{\mu\nu\alpha\beta}$ is the Weyl tensor and ${\mathcal{G}}$ is the Gauss-Bonnet invariant. An additional term $\Box \phi^2$ has to be added to the scalar sector. Let us quote the one loop RGEs for $\lambda$ and $\xi$ ($t\equiv \ln(\mu/M_Z)$) \cite{Bezrukov:2007ep}, :
\begin{eqnarray}
 16\pi^2\frac{d\lambda}{dt} &=& 24\lambda^2 + 12\lambda y_t^2 - 9\lambda(g^2+\frac{1}{3}g'^2)+ .... \nonumber \\
16\pi^2\frac{d\xi}{dt} &=& \left(\xi+\frac{1}{6}\right)[12\lambda + .....]
\end{eqnarray}
where ellipses represent terms proportional to Yukawa and gauge couplings alone contributing to the beta functions and are not relevant for the discussion here. Clearly, in principle, it is possible that $\lambda$ runs to a small value and so does $\xi$. In particular, if $\xi$ is close to the conformal coupling value, it'll stay close to that value even after running while $\lambda$ flows to a very small value, such that an overall consistency with the power spectrum constraints is obtained. There is however one problem though. $\lambda$ should remain fairly constant i.e. should not run in any significant manner for a sufficiently long interval such that enough e-folds are possible for successful inflation. We believe that this is known to experts but it is important to point it out again.

Since the gravitational field is treated as an external field, the quartic coupling beta function does not involve any dependence on $\xi$. One could attempt a better treatment by trying to look at the linearized gravity and considering perturbative graviton corrections. Qualitatively, this will also mean extra bosonic contribution to the quartic coupling which may help soften the vacuum stability problem to some extent. Such a perturbative treatment, though simpler, may be challenged on the grounds of renormalizability. Weinberg had suggested long ago \cite{weinberg-asymp} that gravity may have a fixed point in the ultraviolet (UV). This is referred to as asymptotic gravity or asymptotically safe gravity. There has been a lot of activity in this direction lately (see \cite{Lauscher:2001cq} for an overview). The method relies on non-perturbative exact renormalization group (EREG). Theories with gravity coupled to matter have also been investigated \cite{Percacci:2003jz}. It is generally found that it is quite conceivable that there exist non-trivial fixed points in the UV. We may therefore expect that full SM coupled to gravity will have UV fixed point. Weinberg \cite{Weinberg:2009wa} discussed the possibility of viable inflation within the asymptotic gravity context. In \cite{Tye:2010an} it was pointed out that the scale of inflation and the scale at which the fixed point exists are not the same and there may be issues regarding the number of e-folds in such a situation. It was also suggested that coupling with matter may lead to some differences. In view of all these observations and evidences in favour of a fixed point for various toy theories considered, we conjecture that SM plus gravity will exhibit a fixed point in UV. We may further hope that the flow of the couplings towards the fixed point is rather slow/modest between the GUT scale and the Planck scale. If this does happen to be the case, then Higgs inflation with a small $\xi$ will become a naturally viable scenario, completely consistent with the constraints from CMB power spectrum and spectral index. A small value of $\xi$ also means that the unitarity problem related to Higgs inflation models may no longer be an issue to worry about.
  
Till now the discussion has been around a preferred value of $\xi$ and observational constraints. Let us ask if $\xi$ is needed at all in the first place. As mentioned above, there are reasons to believe that a non-zero $\xi$ is required. Most notably, in the context of quantum theory, $\xi$ gets introduced due to quantum effects and finiteness of the matrix elements of the energy momentum tensor tends to fix $\xi$, depending on the potential and other interactions involved. Let us therefore accept that a non-zero $\xi$ is indeed imposed upon us. This defines the Jordan frame action. The usual strategy employed then is to make a conformal transformation to change frames and go to a description where there is no trace of the non-minimal coupling. At this stage, one may again invoke the restrictions and consistency conditions imposed by the quantum theory. Although a change of variables has an effect of getting rid of the non-minimal term, the theory expressed in terms of the redefined variables is still going to be quantized. After all, the power spectrum or any other correlation function will be calculated employing the quantum fluctuations of the scalar field about the background value. We can not, therefore, just stop at calculating the tree level correlation function, but in the redefined theory must ask exactly the same questions about the renormalization and finiteness of the matrix elements of the energy momentum tensor, now defined in terms of the new field variable. Once such questions are asked, a non-minimal term will become unavoidable, and depending upon the exact form of the potential in terms of the redefined field variable, a value or range of $\xi$ will emerge. We are therefore led to the following general conclusion: {\it in the end, conformal transformations may not eliminate the non-minimal coupling altogether when considering a quantum theory}. Exceptions may exist when the potential in the redefined variables assumes a rather special or completely trivial form.

In passing let us also mention that a very interesting way of generating scale invariant cosmological perturbations was proposed in \cite{Rubakov:2009np} by conformally coupling the scalar field with a negative quartic potential. Somewhat different and more general mechanism is suggested in \cite{Hinterbichler:2011qk}, wherein the field is minimally coupled to gravity and negative quartic potential is employed as a toy example of a rather general class of models. In either case, the basic idea (not getting into the differences between the two approaches) is that the scalar field rolls down the negative potential rapidly and ends up generating scale invariant spectrum. Since the Higgs quartic coupling is expected to turn negative at some point, the Higgs could play the role of the relevant scalar field in such scenarios. Again, a conformal coupling is in accord with the above observations. An attempt to achieve this in the context of inert two Higgs doublet model has been made in \cite{Das:2011wm} by coupling the Higgs fields conformally to gravity.

In this article we have revisited the non-minimal coupling of the inflaton field, with the aim of identifying arguments that may help in pin pointing the value or range of the non-minimal coupling $\xi$. Some of the arguments have appeared before in the literature. We have argued that invoking the conditions of renormalizability and finiteness of the matrix elements of the energy momentum tensor for a given potential may help in fixing $\xi$. Specifically, for $\phi^4$ potential, this leads to $\xi=1/6$, a value that is also preferred if equivalence principle is invoked. It may seem surprising to note that in the leading order in slow roll parameters, imposing the constraints derived from $n_s$ leads to a value of $\xi$ which is not too different from $1/6$. We believe that this is not a mere coincidence and there is perhaps a much deeper reason for this. A large non-minimal term is required to have the correct value of the CMB power spectrum if the quartic coupling has to have its natural value i.e. $\lambda \sim {\mathcal{O}}(0.1)$. This is what is needed in the Higgs inflation models. However, as we have argued, the Higgs quartic coupling in SM  tends to a very small value at the typical scale of inflation, if not negative signaling vacuum instability. Further, we have pointed out that SM coupled to full SM may exhibit a non-trivial fixed point in the ultra-violet. If this does turn out to be correct, it will not only lead to a viable Higgs inflation model but also ameliorate vacuum stability problem. 
It is also worth mentioning that when the scalar field considered does not have a particle physics origin then the RGE governing the running of its quartic coupling will be different than what it is in say SM. In that case, some changes can be expected.
Based on quantum field theoretic arguments,  we also reach a somewhat strong conclusion that conformal transformations may not eliminate the non-minimal coupling completely except in cases when the potential in expressed in terms 
of  the redefined variables assumes a very specific or a trivial form. This, we believe, will have implications for a wide class of models of inflation, and constitutes a key result of the present article.

\end{document}